\documentclass[prc,final,notitlepage,twocolumn,nobibnotes,nofootinbib,superscriptaddress,noshowpacs,centertags]{revtex4}
\usepackage{graphicx,amsmath}
\def\be{\begin{eqnarray}}
\def\ee{\end{eqnarray}}
\def\gsim{\stackrel{\scriptstyle >}{\phantom{}_{\sim}}}
\begin{document}
%=====================================================
\title{CATALYTIC REACTIONS IN HEAVY-ION COLLISIONS}
%=====================================================
\author{\firstname{E.~E.} \surname{Kolomeitsev}}
\affiliation{Matej Bel  University, Bansk\'a Bystrica, Slovakia}%
\author{\firstname{B.} \surname{Tom\'a\v{s}ik}}%
\affiliation{Matej Bel University, Bansk\'a Bystrica, Slovakia}%
\affiliation{Czech Technical University in Prague, FNSPE, Prague, Czech Republic}
%=====================================================
%\date{\today}

\begin{abstract}
We discuss a new type of reactions of a $\phi$ meson production on hyperons, $\pi Y\to\phi Y$ and
anti-kaons $\bar{K}N\to \phi Y$. These reactions are not suppressed according to Okubo-Zweig-Iizuka
rule and can be a new efficient source of $\phi$ mesons in a nucleus-nucleus collision. We discuss
how these reactions can affect the centrality dependence and the rapidity distributions of the
$\phi$ yield.
\end{abstract}
\maketitle
%%%%%%%%%%%%%%%%%%%%%%%%%%%%%%%%%%%%%%%%%%%%%%%
%%%%%%%%%%%%%%%%%%%%%%%%%%%%%%%%%%%%%%%%%%%%%%%
%\section{Introduction}

The study of $\phi$ meson production in different nucleus-nucleus collisions at various collision
energies provides complementary information on collision dynamics and in particular on the
evolution of the strange subsystem. Within the consitutent quark model the $\phi$ meson is
dominantly a spin-one bound state of $s$ and $\bar{s}$ quarks. Hence, hadronic interactions of
$\phi$ mesons are suppressed due to the Okubo-Zweig-Iizuka rule, which, in the strict
implementation, would forbid an interaction of a pure $(\bar{s} s)$ state with non-strange
hadrons. Indeed, the OZI-forbidden reactions are typically orders of magnitude smaller than the
OZI-allowed ones.

Since the OZI suppression weakens the $\phi$ production only by the ordinary hadronic matter and
is lifted in the quark-gluon medium, the strong, order of magnitude, enhancement of the $\phi$ yield was proposed
in~\cite{Shor85} as a signal of the quark-gluon plasma formation. An enhancement of the $\phi$
yield was indeed observed experimentally at AGS and SPS energies albeit to a lesser
degree~\cite{AGS-Back04,NA49-Alt08}. In~\cite{Ko91} it was suggested that the main contribution to
the $\phi$ yields at these energies would be given by the OZI-allowed process with {\it
strangeness coalescence} $K\bar K\to \phi\rho$ and $K\Lambda\to \phi N$\,.

Surprisingly strong enhancement of the $\phi$ yield was observed at the beam energies about
2~GeV per nucleon~\cite{FOPI-Mangiarotti03,Agakichev09}. Such a large $\phi$ abundance cannot be
described by the transport model~\cite{Kampfer02} where $\phi$s are produced in reactions $BB\to
BB\phi$ and $\pi B\to \phi B$ ($B= N,\,\Delta$) with the dominant contribution from pion-nucleon
reactions.
Note that the strangeness coalescence process could not contribute much to the
$\phi$ yield at these energies since kaons have a long mean free path and most likely leave the
fireball right after they are created.

In~\cite{KB09} we propose a new mechanism of the $\phi$ production---{\it the catalytic reactions}
on strange particles
%-------------
\be
\pi Y\to \phi Y\,, \qquad \bar{K} N\to \phi Y\,, \qquad Y=\Lambda\,,\, \Sigma\,.
\label{Cathprod}
\ee
%------------
In contrast to the strangeness coalescence reaction, here the strangeness does not hide inside the
$\phi$s, but stays in the system and the presence of $K$ mesons is unnecessary. The efficiency of
these reactions should be compared with the process $\pi N\to \phi N$, which is found to be
dominating in~\cite{Chung97}. The reactions (\ref{Cathprod}) are OZI allowed, so we win in cross
sections compared to $\pi N\to \phi N$. We lose, however, in the smaller concentration of hyperons
and anti-kaons compared to nucleons and pions.

%%%%%%%%%%%%%%%%%%%%%%%%%%%%%%%%%%%%%%%%%%%%%%%%%%%%%%
\begin{figure}
\centerline{\includegraphics[width=6cm]{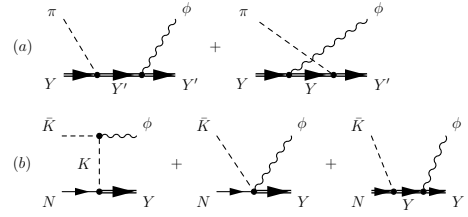}} \caption{Diagrams contributing
to: a)   $\pi\,Y\to \phi \, Y $ reactions; b) $\bar{K} N\to \phi \, Y $ reactions.}
\label{fig:diag}
\end{figure}
%%%%%%%%%%%%%%%%%%%%%%%%%%%%%%%%%%%%%%%%%%%%%%%%%%%%%%

The cross sections of reactions~(\ref{Cathprod}) were calculated
in~\cite{KB09} according to tree-level diagrams (see Fig.~\ref{fig:diag}) given by the
effective Lagrangian of nucleons, hyperons and kaons, which incorporates the $\phi$ meson as a
heavy Yang-Mills boson. Then, one coupling constant fixed by the $\phi\to K\bar{K}$ decay
determines the coupling of $\phi$ to any strange hadron. The resulting isospin-averaged cross
sections can be parameterized as
%-------------
\begin{align}
\sigma_{\pi Y\to\phi Y'}(s)&=p_{\phi Y'}(s)\Big(a_{YY'}+b_{YY'}\frac{\Delta s^{1/2}}{{\rm
1~GeV}}\Big)\, \frac{{\rm mb}}{{\rm GeV}},
\nonumber\\
\sigma_{\bar{K}N\to\phi Y}&=p_{\phi Y}(s)\frac{a_{KY} + b_{KY}\frac{\Delta s^{1/2}}{{\rm 1~GeV}}}
{1+ d_{KY}\frac{\Delta s^{1/2}}{{\rm 1~GeV}}}\, \frac{{\rm mb}}{{\rm GeV}},
\nonumber
\end{align}
%-------------
where $p_{\phi Y}(s)$ is center-of-mass momentum of the $\phi$ meson and hyperon.
For the hyperon channels we have: $a_{\Lambda\Sigma}=4.24$, $b_{\Lambda\Sigma}=1.66$; $a_{\Sigma\Sigma}=2.16$,
$b_{\Sigma\Sigma}=0.851$; $a_{\Sigma\Lambda}=1.40$, $b_{\Sigma\Lambda}=0.682$;
and for the kaon channels:
$a_{K\Lambda}\approx a_{K\Sigma}=2.6$, $b_{K\Lambda}\approx b_{K\Sigma}=1.2$, and
$d_{K\Lambda}\approx d_{K\Sigma}=2.0$

%\section{Strangeness production}

In order to estimate the efficiency of reactions with the $\phi$ production on hyperons and kaons,
we model, first, the evolution of the strange subsystem in the course of a collision. We
make two assumptions: (a) strangeness can be considered perturbatively; (b) the fireball
matter is baryon-dominated. The first assumption allows us to introduce an effective
parameterization for the time dependence of the fireball temperature and baryon density. We use
the form of a scaling solution of hydrodynamic equations
$T(t)/T_m=(\rho_B(t)/\rho_m)^{2/3}=(t^2/t^2_0+1)^{-\alpha}$\,, where $T_m$ and $\rho_m$ are the
maximal temperature and density of the fireball, and $t_0$ is the typical time scale of
the fireball expansion. In the baryon-dominated matter the particles carrying strange quarks,
$\bar K=(K^-,\bar{K}^0)$\,, $\Lambda$\,, $\Sigma$ and heavier hyperons have short mean free path
and remain in thermal equilibrium with pions, nucleons and $\Delta$s till the fireball breaks up.
So, the density of a hadron of type $i$ with mass $m_i$ is equal to
%--------
%\be
$
\rho_i(t) = \lambda_S^{s_{i}}\, \zeta_i\,
e^{q_i\frac{\mu_B(t)}{T(t)}}\, m^2\, T\, K_2\big({m}/{T}\big)/2\pi^2,
$
%\nonumber %\label{h-density}
%\ee
%--------
where $\lambda_S$ is the strangeness fugacity and $s_{i}$ is the number of strange quarks in the
hadron; $\mu_B$ is the baryon chemical potential and $q_i$ is the baryon charge of the hadron. The
degeneracy factor $\zeta_i$ is determined by the hadron's spin $I_i$ and isospin $T_i$ as
$\zeta_i=(2\, I_i+1)\,(2\, T_i+1)$\,. We disregard the finite width of $\Delta$s and treat them as
stable particles with the mass $m_\Delta=1232$~MeV. The baryon chemical potential $\mu_B(t)$
is determined by the equation $\rho_N(t)+\rho_\Delta(t)=\rho_B(t)$\,, where we neglect the
contributions of hyperons, heavier resonances and anti-particles. The strangeness fugacity
follows from the strangeness conservation relation:
$\rho_{\bar{K}}(t)+\rho_\Lambda(t)+\rho_\Sigma(t)\approx \rho_K(t)$, where
$\rho_K(t)$ is the number of produced kaons $K=(K^+, K^0)$ divided by the fireball volume.
We do not assume that $K$ mesons are in thermal in chemical equlibrium with other species
since they have large mean free paths and can leave the fireball at some earlier stage of the
collision.
The evolution of $\rho_K$ is described by the differential equation
$\dot\rho_{K}(t)-\rho_K(t)\, \frac{\dot\rho_B(t)}{\rho_B(t)} =\mathcal{R}(T(t),\rho_B(t))$
with the initial condition $\rho_K(0)=0 $\,.
The kaon production rate, $\mathcal{R}$
is determined by the processes with $\pi N$, $\pi \Delta$, $NN$, $\pi\pi$ and  $N\Delta$
in the initial states; for the list of possible reaction channels and the corresponding cross
sections see Ref.~\cite{Tomasik05}.
%
%%%%%%%%%%%%%%%%%%%%%%%%%%%%%%%%%%%%%%%%%%%%%%%%%%%%%%%%%%%%%%%%%%
\begin{figure}
\centerline{
\parbox{4.1cm}{\includegraphics[width=4.1cm]{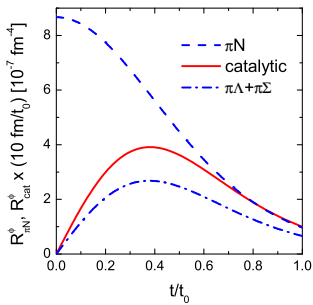}}
} \caption{$\phi$ meson production rates: solid
line is the sum of all catalytic reactions, dash-dotted line the reactions on hyperons.
Dashed line is the rate of the $\pi N\to \phi N$ reaction} \label{fig:FRates}
\end{figure}
%%%%%%%%%%%%%%%%%%%%%%%%%%%%%%%%%%%%%%%%%%%%%%%%%%%%%%%%%%%%%%%%%%
%
%

Now we are in position to analyze whether the catalytic reaction can be an efficient source of
$\phi$ mesons compared to conventional ones. As a baseline we take the $\phi$ production rates in
the $\pi N\to \phi N$ reactions $R^\phi_{\pi N}(t) = \kappa_{\pi N}^{\phi N}\, \rho_\pi\,
\rho_N\,$. The transport coefficient $\kappa_{ab}^{cd}={\langle \sigma_{ab}^{cd} v_{ab}
\rangle}/{(1+\delta_{ab})}$ is the cross section $\sigma_{ab}^{cd}(s)$ of the binary reaction
$a+b\to c + d$ averaged over the momentum distributions of colliding particles with the particle
relative velocity $v_{ab}$. We compare $R^\phi_{\pi N}$ with the catalytic reactions on hyperons
$
R_{\pi Y}^\phi=\sum_{\bar{Y},Y=\Lambda,\Sigma} \kappa_{\pi Y}^{\phi \bar Y}\, \rho_\pi\, \rho_{Y}
$
and on anti-kaons
$
R^\phi_{\bar KN}(t) = \big(\kappa_{\bar{K} N}^{\phi \Lambda}+\kappa_{\bar{K} N}^{\phi \Sigma}\big)\,
\rho_{\bar K}\, \rho_{N}\,.
$
The rates of various processes are shown in Fig.~\ref{fig:FRates} for the maximal temperature and density
$
T_m=130~{\rm MeV}\,,\,\, \rho_m=5~\rho_0\,, \,\, \alpha=0.3\,,
$
where $\rho_0=0.17$\,fm$^{-3}$ is the nuclear saturation density. These parameters correspond
roughly to a collision at beam energy $6A$~GeV. The $\phi$ production in $\pi N$ collisions
(dashed line) starts, of course, at the very beginning and gradually falls off as the fireball
cools down and expands. The rates of catalytic reactions increase initially (solid line) as more strange
particles are produced, reach the maximum at (0.3--0.4)$\,t_0$ and drop off later. The rates
become comparable for times $\gsim 0.6\, t_0$. Note that the dominant contribution is given by
reactions on hyperons (dash-dotted line). The rates in Fig.~\ref{fig:FRates} correspond
to the fireball expansion time $t_0=10$~fm. If the collision lasts longer then the curves for
catalytic reactions have to be scaled up by a factor $t_0/(10~{\rm fm})$, since the number of the
strange particles is proportional to the expansion time. This will make the catalytic reactions
efficient even at smaller temperatures. Our estimates show that the catalytic processes can
contribute the $\phi$ production in heavy-ion collisions.

%%%%%%%%%%%%%%%%%%%%%%%%%%%%%%%%%%%%%%%%%%%%%%%%%%%%%%%%%%%%%%%%%%%%
\begin{figure*}
\centerline{
\parbox{4.1cm}{\includegraphics[width=4.1cm]{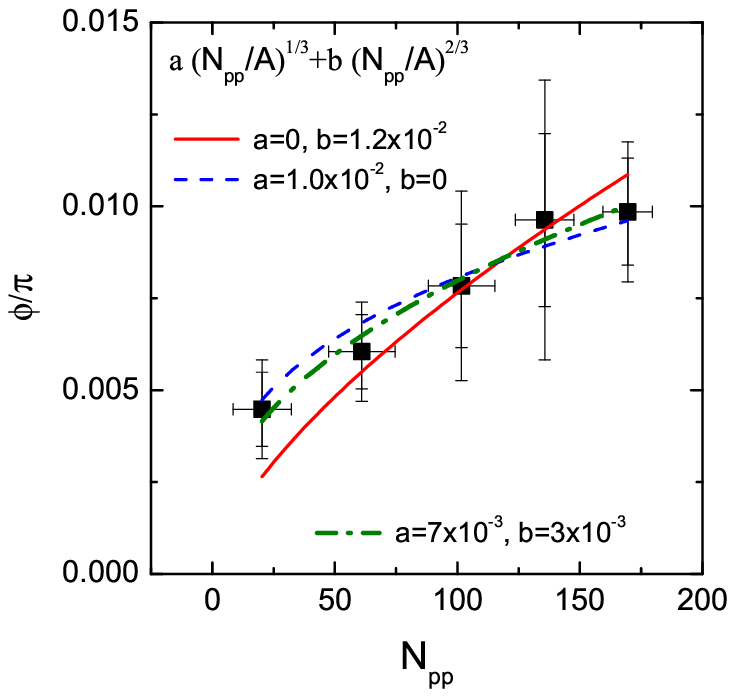}}
\parbox{4.1cm}{\includegraphics[width=4.1cm]{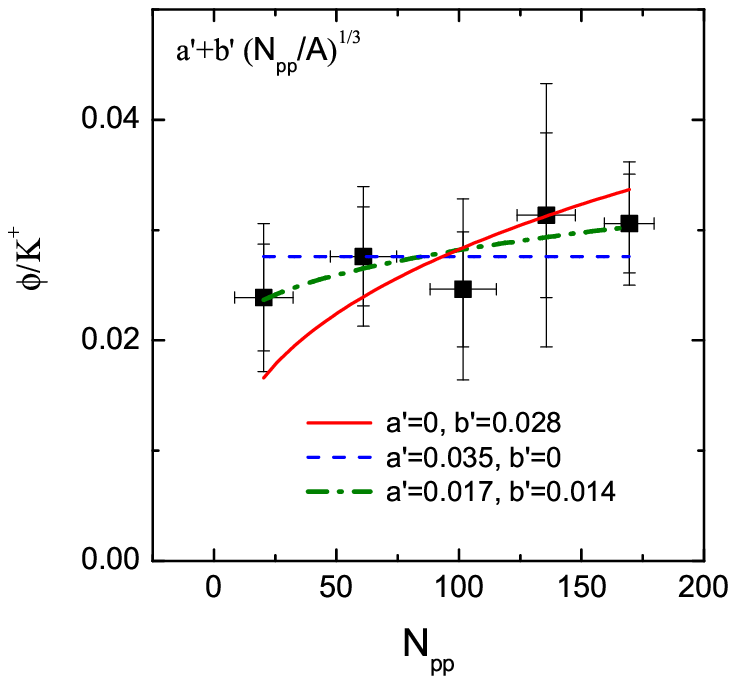}}
\parbox{4cm}{\includegraphics[width=4cm]{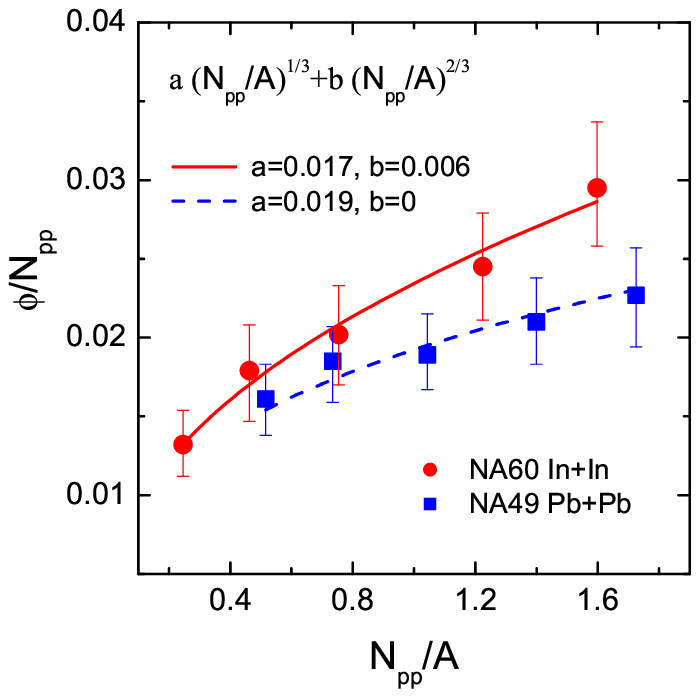}}
}
\caption{Centrality dependence of ratios $\phi/\pi$ (left panel) and $\phi/K^+$ (middle panel).
Data points are for Au+Au collisions at 11.7~$A$GeV~\cite{AGS-Back04}.
Right panel: the $\phi/N_{pp}$ ratio for In+In and Pb+Pb collisions at 158$A$~GeV, the data are from~\cite{In158}.
}
\label{fig:Central}
\end{figure*}
%%%%%%%%%%%%%%%%%%%%%%%%%%%%%%%%%%%%%%%%%%%%%%%%%%%%%%%%%%%%%%%%%%%%

%\subsection{Centrality dependence}

We discuss now the centrality dependence of the $\phi$ production. We will use the mean number of
projectile participants, $N_{\rm pp}$, as a measure for the initial volume of the fireball created in
the collision, $V\propto N_{\rm pp}$. If there is only one changing parameter with the unit of
length as in the case of a symmetrical collision at the fixed collision energy, $l\sim
V^{1/3}\propto N_{\rm pp}^{1/3}$, the scaling properties of hydrodynamics imply that the collision
time is of the order $t_0\sim l/c\propto N_{\rm pp}^{1/3}$~\cite{Russkikh92}. Then the number of
produced $\phi$ mesons can be estimated as $N_\phi \sim a_{\rm conv}\,N_{\rm pp}^{4/3} + a_{\rm
cat} \,N_{\rm pp}^{5/3}$. The term proportional to $N^{4/3}_{\rm pp}$ is due to the conventional production
reactions like $\pi N\to\phi N$, whereas the term proportional to  $N^{5/3}_{\rm pp}$ corresponds to the
catalytic reactions~\cite{KB09}. For the experimental ratios in~\cite{AGS-Back04} we find
$
{N_\phi}/{N_\pi}\sim a\,n_{\rm pp}^{1/3}+b\, n_{\rm pp}^{2/3}$\,,
and
${N_\phi}/{N_{K^+}}\sim a' + b'\, n_{\rm pp}^{1/3}\,,
$
where $n_{\rm pp}=N_{\rm pp}/A$, $a\,,a'$ and $b\,,b'$ parameterize the relative strength of conventional and
catalytic processes, and $A$ is the number of nucleons in the colliding nuclei.
In Fig.~\ref{fig:Central} we compare these parameterizations with the
available data for Au+Au collisions at
11.7~$A$GeV~\cite{AGS-Back04}. First, we adjust parameters $a,a'$ and $b,b'$
separately and obtain dashed and solid curves, respectively.
The optimal
fits are reached when both parameters are activated, dashed-dotted lines in
Fig.~\ref{fig:Central}.
On the right panel in Fig.~\ref{fig:Central} we show the fit of the $\phi/N_{\rm pp}$ ratio
for In+In and Pb+Pb collisions at 158$A$~GeV~\cite{In158}. The data of In+In collisions
can be better described with the account for the catalytic reactions.

%\subsection{$\phi$ rapidity distribution}

%%%%%%%%%%%%%%%%%%%%%%%%%%%%%%%%%%%%%%%%%%%%%%%%%%%%%%%%%%%%%%%%%%%%
\begin{figure}
\centerline{
\parbox{4.1cm}{\includegraphics[width=4.1cm]{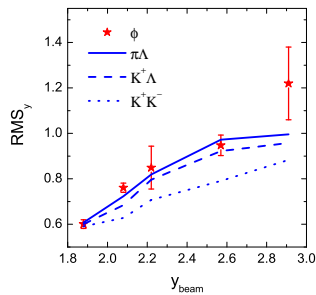}}
}
\caption{Root mean square of the rapidity distributions of $\phi$s produced in Pb+Pb
collisions versus the beam rapidity~\cite{NA49-Alt08,In158}. Lines show the distribution widths from reactions
$\pi\Lambda\to \phi Y$, $K^+\Lambda\to \phi N$ and $K^+K^-\to \phi$. }
\label{fig:Rap}
\end{figure}
%%%%%%%%%%%%%%%%%%%%%%%%%%%%%%%%%%%%%%%%%%%%%%%%%%%%%%%%%%%%%%%%%%%%

The systematics of $\phi$ rapidity distributions in Pb+Pb collisions at the SPS is reported
in~\cite{NA49-Alt08}. The distributions are given by the sum of two Gaussian functions of width
$\sigma$ placed symmetrically around mid-rapidity $y_{\rm beam}$ at distance $a$. The width of the
distribution is characterized by the root mean square ${\rm RMS}^2=\sigma^2+a^2$\,.
Ref.~\cite{NA49-Alt08} pointed out that the width of the $\phi$ meson distribution does not fit
into the systematics for other mesons, increasing much faster with $y_{\rm beam}$. It was
emphasized in~\cite{NA49-Alt08} that the steep rise of the $\phi$ distribution width cannot be
explained by the hadronic process $K^+\, K^-\to \phi$. We note that the rapidity distribution of
hyperons increases much faster than those for mesons as the hyperons are dragged with nucleons to
forward and backward rapidities. We assume now that the rapidity distributions of particles do
not change after some initial stage when nuclei are passing through each other. The reaction
kinematics is restricted mainly to the exchange of transverse momenta. Then the rapidity
distribution of $\phi$s produced in the reaction $1+2\to \phi +X$  is roughly proportional to the
product of rapidity distributions of colliding particle species 1 and 2. Using the parameters of
$K^+$ and $K^-$ from~\cite{NA49-Afanasiev02,NA49-Alt08-piK}  we obtain {\rm RMS}s for the  $K^+\,
K^-\to\phi$ reaction shown in Fig.~\ref{fig:Rap} by dotted line.  In contrast, the width of $\phi$
rapidity distributions from the reactions involving $\Lambda$ particles,  $\pi\Lambda\to \phi Y$
and $K^+\Lambda\to \phi N$ (solid and dashed lines) rises much faster and is comparable with the
experimental results.

%\section{Conclusions}
In conclusion, we proposed a new mechanism of $\phi$ meson production in nucleus-nucleus
collisions---the catalytic reactions on strange particles (\ref{Cathprod}). These reactions are
OZI-allowed and their cross sections can be by an order of magnitude larger than the cross
sections of conventional OZI-suppressed $\phi$ production reactions $\pi N\to \phi N$ and $N\Delta\to\phi N
$, etc. We showed that the rates of $\phi$ production in the catalytic reactions can be competitive
to or even exceed the conventional reactions. The catalytic reaction could affect the centrality
dependence of the $\phi$ yield. Analyzing the $\phi$ rapidity distributions at SPS energies we
find that the strong rise of the distribution width with the collision energy can be explained by
the $\phi$ production in $\pi\Lambda$ and  $K^+\Lambda$ collisions.

We gratefully acknowledge the support by the VEGA grant 1-0261-11 of the Slovak Ministry of Education.

\end{document}